\def\year{2020}\relax
\documentclass[letterpaper]{article} 
\usepackage{aaai20}  
\usepackage{times}  
\usepackage{helvet} 
\usepackage{courier}  
\usepackage[hyphens]{url}  
\usepackage{graphicx} 

\usepackage{xcolor}
\usepackage{multirow}
\usepackage{amsmath}
\usepackage{booktabs}
\usepackage[normalem]{ulem}
\usepackage{subcaption,dcolumn,adjustbox,siunitx}

\usepackage{graphicx}  
\begin{document}
%

\setcounter{secnumdepth}{0} 

%
\setlength\titlebox{2.5in} 

\title{Towards Measuring Adversarial Twitter Interactions against\\ Candidates in the US Midterm Elections}
\author{Yiqing Hua, Thomas Ristenpart, Mor Naaman\\ 
Cornell Tech\\ 
2 West Loop Road\\
New York City, New York 10044\\
yiqing@cs.cornell.edu, ristenpart@cornell.edu, mor.naaman@cornell.edu 
}

\maketitle
\begin{abstract}
Adversarial interactions against politicians on social media
such as Twitter have significant impact on society.
In particular 
they disrupt substantive political
discussions online, and may discourage people from seeking public office.
In this study, we measure the adversarial interactions
against candidates for the US House of Representatives during the run-up to the 2018 US general election. 
We gather a new dataset consisting of 1.7 million tweets involving candidates,
one of the largest corpora focusing on political discourse.
We then develop a new technique for detecting tweets with toxic content that are directed at any specific candidate.
Such technique allows us to more accurately quantify adversarial interactions towards political candidates. 
Further, we introduce an algorithm to induce
candidate-specific adversarial terms to capture more nuanced adversarial
interactions that previous techniques may not consider toxic.  Finally, we use these techniques to outline the breadth of adversarial interactions seen
in the election, including offensive name-calling, threats of violence, 
posting discrediting information, attacks
on identity, and adversarial message repetition. 
\end{abstract}
\section{Introduction}

The growing trend of incivility in online political discourse has important
societal ramifications~\cite{nytimes,ukreport}. The negative discourse is 
discouraging politicians from engaging in conversations with users on social
media~\cite{theocharis2016bad}, has caused some candidates to drop out of
races~\cite{nytimes}, and has unknown impact in terms of chilling others from engaging in 
democracy.  
Within a large body of recent work on online abuse and harassment, there is increasing interest in understanding and measuring abuse towards political figures~\cite{gorrell2018twits,adversarial}. 

In this work we focus on improving the understanding of online adversarial interactions
in political contexts, using as a case study 
the 2018 midterm elections for all 435 seats in the US House of
Representatives. 
We broadly define adversarial interactions as messages intending to hurt, embarrass,
or humiliate a targeted individual.
Such behaviors include explicitly abusive or harassing language targeted at a candidate 
as well as more implicit actions aiming to discourage or discredit individuals, for example posting  misinformation and subtle personal attacks.
To perform this analysis, we collect a dataset of tweets, retweets, mentions, and replies involving a set of 786 candidates  over several months leading up to the US House election. With 1.7 million tweets, the resulting dataset is one of the largest datasets available of directed social media interactions with political candidates. 

Analyzing adversarial interactions in such a large dataset faces several challenges.
The sheer size of the data requires scalable, automated detection approaches. 
However, detection approaches used in previous measurement studies are often based on either language models trained on a corpus with little social context~\cite{mondal2017measurement,salminen2018anatomy} or explicitly offensive lexicons~\cite{gorrell2018twits}, thereby focusing on the severity of the language used in messages.
As a result, these context-agnostic language-based approaches may miss certain kinds of adversarial actions that don't include severe language. 
Further, these detection techniques~\cite{yin2009detection,nobata2016abusive,wulczyn2017ex}
often assume the toxicity in the interaction is directed towards the receiver. 
For example, previous work~\cite{gorrell2018twits} assumed all Twitter replies with 
toxic content are directed at the account being replied to,
which as shown below may lead to over-counting the amount of abuse received by some individuals.
In this work, we first examine and improve on the precision of automated language-based techniques. We then explore, using a new method, what kinds of adversarial interactions these techniques may overlook. 

Our first goal is to characterize adversarial interactions against political candidates with the above challenges in mind.
We build on existing techniques that provide high precision and scalable toxicity detection to explore the candidate attributes--including gender and affiliated party--that are associated with the amount of adversarial interactions candidates receive.
To help with this task, we use the interaction context to design heuristics in order to infer the direction of toxicity: 
given an utterance and its receiver~(e.g. the account that was replied to or mentioned in a tweet), 
determine if the utterance contains toxic content \textit{directed at the receiver}.
A key insight is that we can leverage the partisan nature of US politics for this analysis.
Specifically, we combine Perspective API~\cite{perspectiveapi}, a state-of-the-art language-based toxicity  detection tool, with heuristics that predict 
political leaning of the users to help determine the likely target of the abuse.
We show that using these heuristics improves the precision of harassment detection directed at candidates compared to using Perspective API alone. 

While the precision is high, the potential downside of using general language models trained in different context is low recall.
In order to examine the limitations of general language models in this specific context (i.e. discourse with political candidates),
we provide a new approach for discovering \emph{target-specific
adversarial lexicons} that uses the political network context and a label propagation algorithm to expose phrases that are likely to be used in an adversarial way against a specific candidate, but are not necessarily abusive in other contexts. 
Using this approach, we provide evidence and examples of adversarial interactions missed by the general language model approach.

In conclusion, we propose techniques that allow better quantification of adversarial interactions towards candidates. 
In addition, we design a novel discovery algorithm that exposes a 
diverse set of ``personalized'' adversarial interactions that are not detected via context-agnostic harassment detection techniques. 
This paper therefore provides new insights into the nature of adversarial
interactions in the important context of online democratic processes. 

\section{Related Work}

\subsubsection{Measuring adversarial interactions.} Most previous measurement
studies on online abuse focus on \textit{generalized} hate speech or abuser
characteristics~\cite{mondal2017measurement,finkelstein2018quantitative,chatzakou2017measuring,chatzakou2017hate,ribeiro2018characterizing,chatzakou2017mean}.
Most similar to our work, Gorrell et
al.~\cite{gorrell2018twits,gorrell2018online} used dictionary-based method to
measure abusive replies towards UK parliament members on Twitter, in order to
understand how quantity of hateful speech is influenced by factors including
candidate gender, popularity etc.  However, the analysis ignores the fact that in
communications on Twitter, the usage of hate words in reply tweets might not
be abusive towards the recipient being replied to.  Different from previous
approach, in our analysis, we define the problem of \textit{directed toxicity}
detection and develop an approach by using both content and user features as the
first attempt to address it.

\subsubsection{Categorizing adversarial interactions.} 
Previous works have worked on characterizing adversarial interactions in order to
design better annotation schemes~\cite{founta2018large}, understand victim's experiences~\cite{matias2015reporting}
or to identify different themes in \textit{generalized} hate speech~\cite{elsherief2018hate}.
Unlike previous work, we focus our analysis on \textit{directed} harassment towards politicians and
develop a framework to identify \textit{target-specific adversarial lexicons}.
With our technique, we are able to discover contextual adversarial topics that are typically missed by existing machine learning techniques.
In previous research, categorization of adversarial interactions
are often coded at comment level~\cite{salminen2018anatomy},
and inevitably ignores certain harassment categories such as sustained harassment towards individuals.
In contrast, we analyze \textit{directed} adversarial interactions at the target level.
To obtain a more exhaustive list of types of adversarial behaviors,
we combine our categorization with typologies from research that examined victim reported harassment~\cite{duggan2014online,matias2015reporting},
and present examples in our dataset from each category.

\section{A Political Interactions Data Set}
\label{sec:data}

\subsubsection{Data collection.}
Our goal is to use a data-driven approach in order to obtain a better understanding of
adversarial interactions with political
candidates online. We retrieved the full list of candidates running in 2018 for the
United States' House of Representatives from
Ballotpedia~\cite{ballotpedia}.
We filtered out candidates who didn't pass the
primary election (except for those in Louisiana, where the
primary election is held with the general election),
resulting in 1-2 candidates for each of the 435 congressional races.

We obtained
the candidates' Twitter accounts by manually verifying the campaign
accounts listed on their Ballotpedia pages and campaign websites. We included
the candidates' personal or office accounts (for incumbents) when found.  Our final
dataset includes a list of 786 candidates ($87\%$ of all House candidates
competing in November, 2018): 431 Democrats (D) and 355 Republicans (R)
candidates from all 50 states with $1,110$ Twitter accounts in total. 
We obtained the gender of each candidate based on manual inspection on candidate profiles. In total, our dataset includes 231 female candidates
and 555 male candidates\footnote{The account database is available at~\url{https://github.com/vegetable68/Midterm-2018-candidates}.}.

We collected data from Twitter~(using the Twitter streaming API) from September
17th, 2018 until November 6th, 2018, including all tweets posted by,
mentioning, replying to, or retweeting any of the candidate accounts. 
We estimate good coverage on all data except mentions 
due to the limited access of the Twitter API.
In total, our data consists of 1.7 million tweets and 6.5 million retweets of candidates from 992 thousand users (including the candidate accounts).
We publish all the tweet ids collected at Figshare~\footnote{\url{https://figshare.com/articles/U_S_Midterm_Election_Twitter_Dataset_2018/11374062}}.

On Twitter, following relationship among users has been shown to be 
informative for inferring user interest or political 
preference~\cite{romero2013interplay,barbera2015birds}.
We therefore retrieved the 5,000 most recently 
followed accounts by each user account via the Twitter Standard API~\cite{twitterapi}. 
Due to Twitter API rate limits,
this data collection was only completed by March 2019.
There are two main limitations of this network data.
First, for users who have left Twitter or set their profiles to be private at the time of the data collection,
their friends lists were not retrieved.
Nevertheless, as we prioritized the more active users in our dataset (users who interacted more with candidates by replying to their tweets or mentioning their accounts) while collecting this data,
we consider these omissions as non-critical.
Second,  as users may follow more than 5000 accounts,
or follow new accounts after their interaction with political candidates was recorded,
our data collection is not entirely accurate. 
In total, we obtained full Twitter friends 
list of $92\%$ of all users in our data as they follow fewer than $5000$ accounts.
Moreover, we retrieved partial friends lists for $7\%$ of all users who
followed more than 5000 accounts.
The remaining $1\%$ accounts were either deleted or
suspended at the time of our network data collection.

Additionally, we performed manual labeling for a subset of user profiles and
tweets in order to verify and improve results of our analysis.
Details of the annotation tasks are introduced when the data is used in the following discussions.
We use a team of three
graduate students and two researchers.  The labeling is done by two raters and
conflicts are resolved by the third.

\subsubsection{Attention.} 
Before turning to analyzing the adversarial interactions, we provide some basic analysis
of interactions in the dataset. An \emph{interaction} with a candidate is a tweet
that is either a reply to a tweet by the candidate, or a tweet that
mentions the candidate.
We define the \emph{attention} received by a candidate to be the number of interactions
towards them in the dataset~(i.e. mentions or replies).
The distribution of attention varies
significantly across candidates. Over our data
collection period, for example, we measured 82,100 replies to the tweets of Nancy
Pelosi, while another candidate, Erika Stotts Pearson, had a total of five.

\begin{figure}[t]
    \begin{subfigure}[b]{0.23\textwidth}
        \centering
        \includegraphics[width=\textwidth]{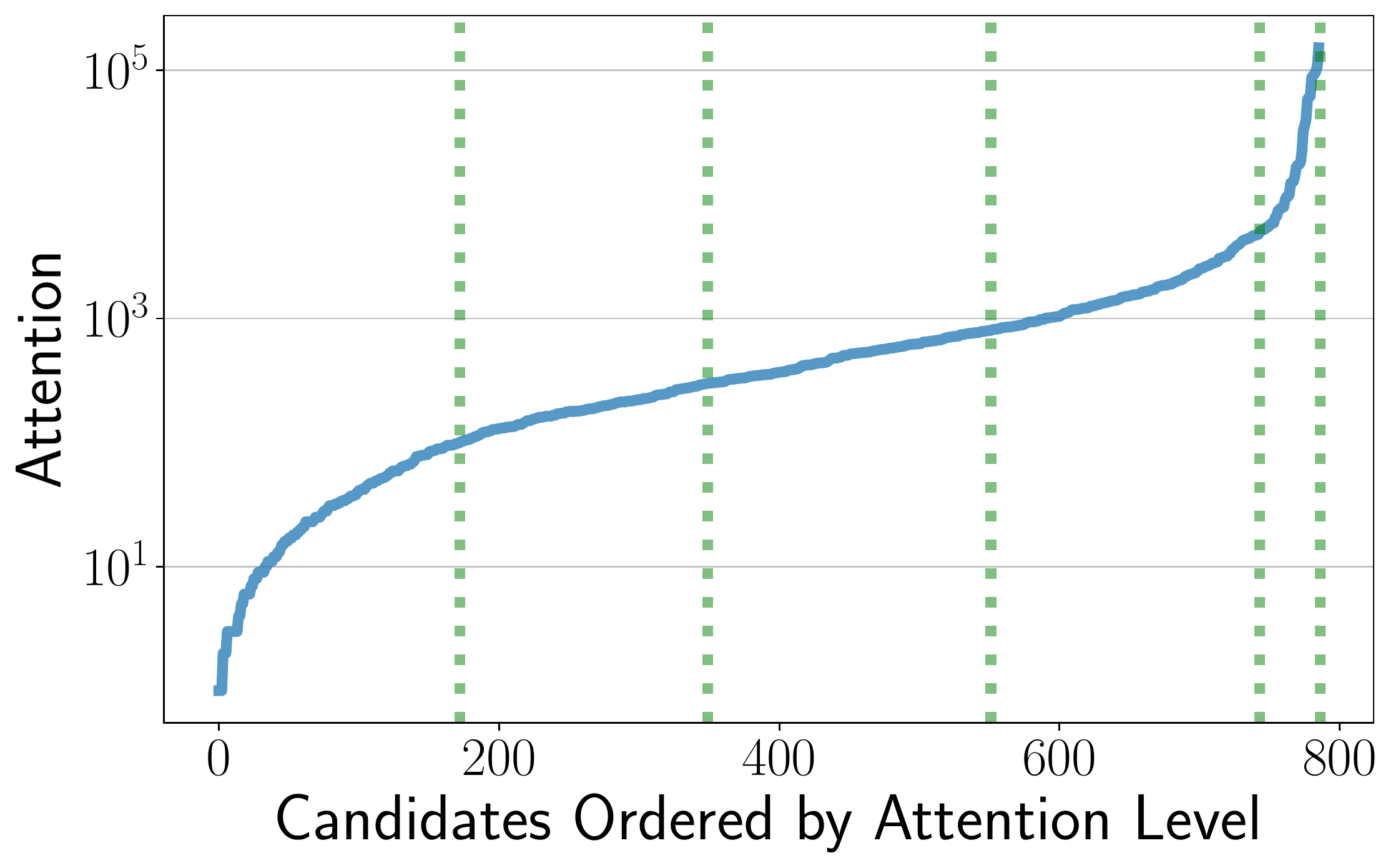}
        \caption{\label{fig:attentiondist1}}
    \end{subfigure}
    \hfill
    \begin{subfigure}[b]{0.23\textwidth}
        \centering
        \includegraphics[width=\textwidth]{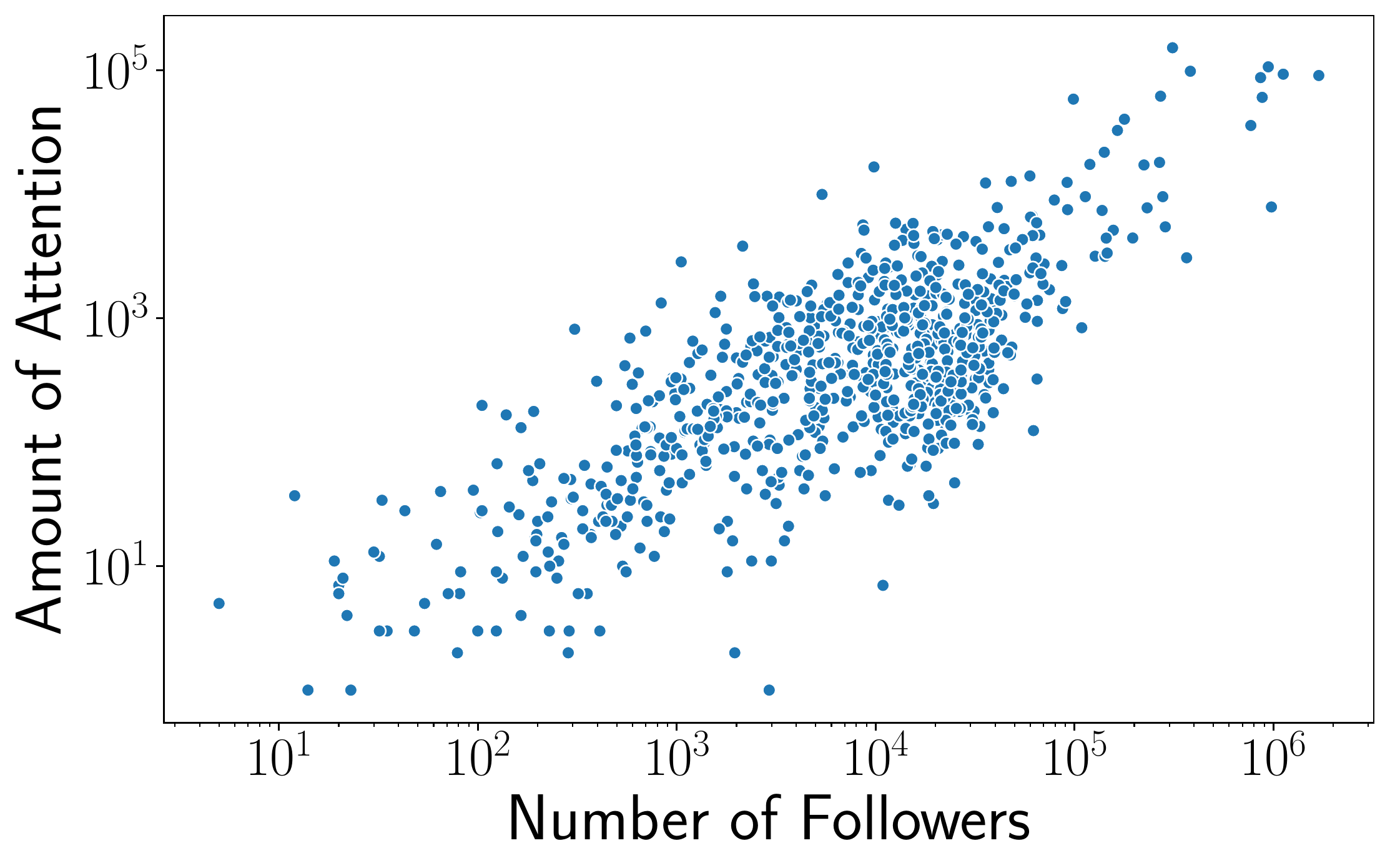}
        \caption{\label{fig:attentiondist2}}
    \end{subfigure}
  \caption{\label{fig:attentiondist}
  (a): Amount of attention received per candidate and attention tiers separated by vertical dotted line.
  (b): There's strong correlation between amount of attention and number of followers per candidate.
  }
\end{figure}

In Figure~\ref{fig:attentiondist}, we show the distribution of attention
received by candidates (left) and the correlation between the attention and the
number of followers each candidate has (right).  Clearly, the $786$ candidates
in our dataset receive different levels of attention and may experience and
perceive adversarial interactions differently, as attention is heavily skewed
towards a few high-profile candidates and has a long tail of candidates without
much attention attracted.  

\section{Directed Toxicity Detection}

\begin{table*}[t]
  \centering
  \footnotesize
  \begin{tabular}{lrr|rrrrr}
  \toprule
                 & \multicolumn{2}{c|}{Any Candidate} & Tier 1 & Tier 2& Tier 3 & Tier 4 & Tier 5\\
                 & {Replies}&{Mentions}               & \multicolumn{5}{c}{(All interactions)}\\
  \midrule
  	  Adversarial tweets (manual labels) & 97  & 94 & 97 & 92 & 95 & 93 & 96 \\
  	  Adversarial \& directed at candidate (manual labels) & 75  & 56 & 70 & 71 & 65 & 68 & 73 \\
  \midrule
      Automatically labeled by DPP & 96 & 89 & 91 & 91 & 95 & 90 & 94\\
  	  Adversarial \& directed in DPP-labeled set (manual labels) & 71  & 52 & 62 & 64 & 62 & 62 & 68 \\      
  \midrule
  DPP precision (DPP-labeled set) & 93\% & 68\% & 77\% & 81\% & 88\% & 84\% & 91\% \\
  DPP recall (DPP-labeled set) & 92\% & 73\% & 85\% & 81\% & 81\% & 82\%  & 90\% \\
  \bottomrule
  \end{tabular}

  \caption{\label{tab:directedSample} Comparison of directed toxicity detection
  approaches. (Top) The number of tweet-candidate pairs manually labeled
  as adversarial in general and adversarial towards the candidate, out of 100 
  tweet-candidate pairs randomly chosen from tweets marked as adversarial by the Perspective API.
   (Middle) The number of tweet-candidate pairs for which we had
    enough information to label via DPP across different categories,
    and the number of adversarial tweets directed at candidates in the
    DPP-labeled sets.
  (Bottom)  The precision and recall of the DPP method on the labeled set.
    }
\end{table*}

Given the scale of our data, we have to rely on methods that can scale for a quantitative analysis.
To ensure the accuracy of such methods, we first define the problem of
detecting \textit{directed toxicity} within tweets. Given a tweet, a set of
possible targets of abuse (recipients of, or mentions within, the tweet),  one can use
both language and social clues, such as online community structure or user
social network, to determine whether the tweet contains adversarial content
towards one or more of the target(s).  In this section, we use our dataset to
show the insufficiency of adversarial interaction detection approaches that lack directionality, and then provide a new method,
\textit{directionality via party preference} (DPP) as the first attempt at
solving it.

\subsubsection{Insufficiency of prior approaches.} 
Previous work~\cite{gorrell2018twits} used a combination of dictionary-based
techniques and Twitter metadata, such as the fact that tweets include mentions
and replies, to infer adversarial interactions. We improve on this along two
dimensions, first replacing dictionary-based techniques with state-of-the-art
machine learning techniques, and, second, providing a more advanced solution
to the directionality problem.  

We use Perspective API~\cite{perspectiveapi} to determine if a tweet contains
adversarial content.  We use the API's TOXICITY score, which indicates whether the
model believes the utterance to be discouraging participation in conversations.
The score is a value in $[0,1]$, with 1 indicating high likelihood of containing toxicity and 0 being
unlikely to be toxic.
In order to choose a threshold with high precision in detecting adversarial tweets, 
we created the following validation set with $100$ tweets.
For each toxicity interval with length $0.1$~(i.e. $(0, 0.1]$, $(0.1, 0.2]$, etc), we sampled $10$ tweets and asked annotators if it contains adversarial content.
Annotators labeled $44\%$ of the tweets as adversarial.
When choosing $0.7$ as the threshold, the detection has the highest precision $90\%$ while maintaining a reasonable recall at $61\%$,
with F1=$0.73$.
Therefore, in the following discussions, 
we choose $0.7$ as the threshold for marking a tweet as adversarial.
A further validation of this threshold on the same dataset shows that results of analysis on adversarial interactions against political candidates are robust to small changes in this threshold~\cite{adversarial}.

One alternative approach to identify toxicity in tweets is to classify the sentiment in their contents.
However, scores assigned by sentiment analysis are not sufficient to reflect the adversarial-ness of tweets.
To experiment, we used vader sentiment analysis package~\cite{hutto2014vader} to assign sentiment scores to the tweets in the above described validation set.
Given a tweet, the algorithm outputs a continuous value between [-1, 1] indicating the sentiment conveyed in the utterance.
We chose the threshold for classifying a tweet as adversarial being the threshold that makes the highest F1 score.
The resulted precision is $66\%$ with recall of $78\%$ and F1 of $0.72$.
We used Perspective API for the following analysis as it has higher precision in detecting adversarial tweets.

\begin{figure}[t]
\center
  \includegraphics[width=0.4\textwidth]{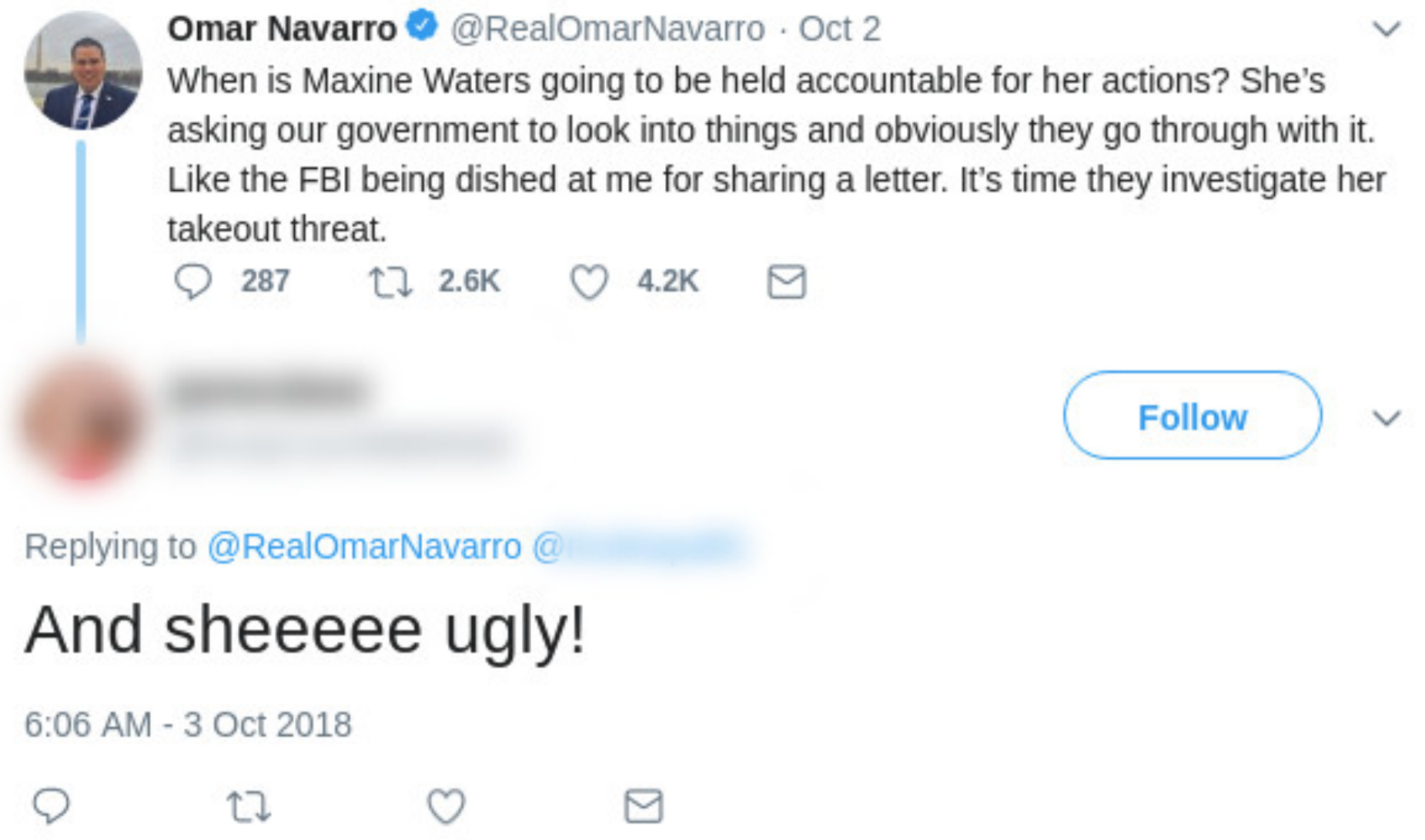}
  \caption{\label{fig:tweet-example}
  An example tweet from a user attacking Maxine Waters while replying to Omar Navarro.}
\end{figure}

That leaves the second challenge, inferring directionality. 
In our data users tend to attack a candidate's opponent while replying to them,
making the prior approach -- simply looking at the Twitter reply or mention metadata -- insufficient.
To make this concrete, consider house candidate Omar Navarro (R),
who was running against longtime house member Maxine Waters (D).
Since Waters has been very vocal in terms of her attitude towards President Trump,
she attracts a large amount of attacks from pro-Republican users, even when they
are replying to Navarro's tweets.
An example of this  
is given in Figure~\ref{fig:tweet-example}.
In a human validation of 100 machine-labeled adversarial tweets replying to Navarro,
we notice that although 91 tweets in total contain adversarial content,
only 18 of them target Navarro. 
In this case, the straightforward combination of Perspective API with 
Twitter metadata would overcount the number of adversarial interactions towards Navarro.

To understand how general the overcounting problem is, we perform an experiment
with tweet-candidate pairs.  We annotate seven separate
sets of 100 tweet-candidate pairs, randomly sampled from tweets labeled as
adversarial by Perspective API.  The first set includes tweets that are replies to any
candidate, and the second set includes tweets that 
mention any candidate.
The last five sets of tweets are ones 
interacting with candidates with different levels of popularity.
For this purpose, we divided candidates into five tiers according to the different amount of attention they received.
Details of the grouping are shown by the dotted lines in Figure~\ref{fig:attentiondist}(a).
Each tweet and candidate pair is labeled with two questions:
(1) does this tweet contain adversarial content; 
(2) is the adversarial content targeting the candidate.

We present the results of the analysis in the upper section in Table~\ref{tab:directedSample}~(The middle and bottom section are explained below.)
The first row shows the number of tweets that indeed contain adversarial content.
On average, Perspective API has a precision of $95\%$ in detecting adversarial content.
The second row shows the number of tweets that are adversarial against the
candidate being replied to or mentioned.
For just $75\%$ of the replies and $56\%$ of the mentions, 
the tweet contains adversarial content targeting the replied-to or mentioned
candidate respectively. 
This precision suggests that simply relying on Twitter metadata is not sufficient in understanding the targets of adversarial content.

\subsubsection{Directionality via party preference.}
We introduce a new set of heuristics for determining if a candidate being mentioned or replied to is the target of adversarial content. 
With the partisan nature of political discussions in the U.S., 
we assume that when a candidate from one party is replied to or mentioned in an
adversarial tweet by a user that displays an affinity for that party,
it is more likely that the hostility is actually towards the opposing candidate
or party. To take advantage of this insight, we need a way to infer a user's political
leaning, which we describe in a moment. Assume for now we can infer a user's
party preference.
Then, our \emph{directionality via party preference (DPP)} method labels a tweet as
adversarial to a candidate if it is machine-labeled as adversarial and the tweet's author leans towards the political party
opposing that of the candidate.

\subsubsection{Inferring user party preference.} We now aim to infer
party preference of users.  To be specific, for each user in our dataset, we
assign a party preference tag of either pro-Democrat, pro-Republican, or
unknown.  The tag is assigned combining three features: hashtags in user
profiles, users' retweet patterns and following relationships 
on Twitter.

\emph{(1) Hashtags in user profile:} We adapt an approach
from previous work~\cite{conover2011political} to bootstrap politically polarized hashtags in
user profiles.  We begin by seeding from two highly politically polarized
hashtags, \texttt{\#maga} (Make America Great Again, typically used by
pro-Republican users) and \texttt{\#bluewave} (typically used by pro-Democrat
users).  Then we identify a set of hashtags that are related to the seeds by
examining co-occurrence in user profiles.  For a set of user profiles $S$
containing a seed hashtag $s$, and a set of user profiles $T$ containing a
non-seed hashtag $h$, the political relatedness of $h$ to seed hashtag $s$ is
assigned to be the Jaccard coefficient between $S$ and $T$, i.e., $|\frac{S \cap
T}{S \cup T}|$.  Using a threshold from previous work, hashtags with political
relatedness larger than $0.005$ are considered related to the seed hashtag.
We populate the $S$ and $T$ sets
with any relevant user from our dataset, and this results in 55 and 64 hashtags related to \texttt{\#maga} and
\texttt{\#bluewave} respectively, with zero overlap between the two sets of
hashtags.  
We manually filter out the hashtags that are ambiguous in terms of representing a
political preference (for example, \texttt{\#israel}).  The full list of
politically related hashtags and the list of hashtags used in labeling user
political preference are both shown in the Appendix.
We then label users as pro-Democrat or pro-Republican by hashtag occurrence
in the user profile.  Users without any hashtags from either list or with
hashtags from both lists are labeled as unknown.  

\emph{(2) Retweet pattern:}
Previous work~\cite{conover2011political} shows that political retweet networks in Twitter are highly
polarized.
We assign a user's party preference by their retweeting behavior during our data collection period.
Users are labeled as pro-Democrat if they retweet more from Democrats than from Republicans,
and labeled as pro-Republican if it is the other way round.
Users with no retweet records or that retweet equally from both parties, are labeled as unknown.

\begin{table*}[t]
    \small
    \begin{tabular}[t]{p{50mm}S[table-format=1.2]@{\hskip 0.4in}r@{\hskip 0.3in}lrr}
      \toprule
      &\multicolumn{2}{c}{\textbf{Regression Model}}&
      \textbf{Distribution}&\textbf{Mean}&\textbf{Standard deviation}\\
      & $B$ & $SE$& \\
      \midrule
      (Intercept) & 44.3*** & 2.65 &\\
      \textit{Amount of attention from \newline users supporting opposing party} & 117.1*** & 7.27 & 
      \begin{minipage}{.2\textwidth}
        \includegraphics[width=\linewidth]{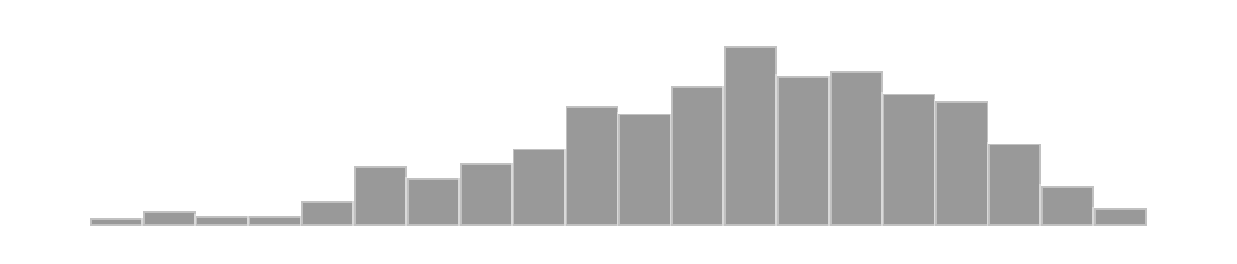}
      \end{minipage} & 2.06 & 0.66\\
      \textit{Number of followers} &  -21.0**  & 7.07 & 
      \begin{minipage}{.2\textwidth}
        \includegraphics[width=\linewidth]{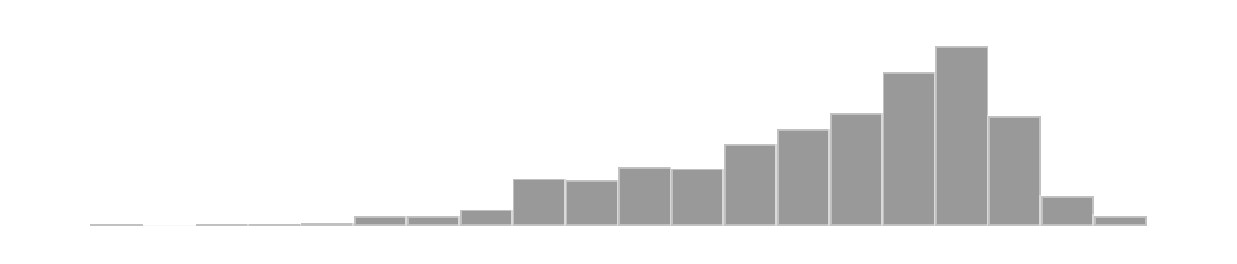}
      \end{minipage} & 3.85 & 0.42\\
      \small{Candidate gender} &  4.09 & 5.91 &
      $30\%$ of the candidates are female & 0.7\\
      \small{Candidate party} & 4.65  & 6.11 & 
      $59\%$ of the candidates are Democrats & 0.41\\
      Candidate gender $\times$ party &  -25.3* & 12.8 && \\
      \midrule
      $R^2$  &\multicolumn{2}{c}{0.388} \\
      \bottomrule
      \multicolumn{6}{l}{Significance codes: *** $p < 0.001$, ** $p < 0.01$, * $p < 0.05$}
      \end{tabular}
      \caption{\label{tab:regression} Results of regression analysis for number
      of adversarial tweets received by a candidate ($N=707$).
      Log transformed variables are listed in italic with mean and standard deviation in log scale.
      The histograms indicate the data distribution of variables after standardizing.}
\end{table*} 

\emph{(3) Friendship on Twitter:} 
Previous work~\cite{barbera2015birds} used features from a user's following relationships 
to successfully predict the user's political preference.
Here we use the same resource for the same purpose.
First, we connect user pairs with bidirectional edges if one of the two follows the
other. 
In our dataset, we observe that the friendship network among users 
is separated into two communities.
Therefore,
using a label propagation algorithm~\cite{raghavan2007near} that assigns community labels in a network,
from a seed set of users labeled by the two previously introduced approaches (excluding users with disagreeing labels),
we iteratively update each user's political preference label according to the label that is shared by most of its neighbors.
Ties are broken randomly when they occur. 

Finally, a user's political label is assigned by the majority vote among the three methods.
In total we are able to label $98\%$ of all
users with a political leaning label. 
Furthermore, for users who are relatively more active in terms of
interacting with candidates, it is more likely that we have
enough information to label their political preference. 

Note that it is likely that some of the users captured in our dataset are automated bots~\cite{davis2016botornot,ferrara2016rise},
or part of state-operated information operations~\cite{im2019still,arif2018acting}.
However, such activity is of minimal impact on our result,
as adversarial interactions that are generated by bots or humans are both perceived by candidates and other potential audience,
with the same negative impact.
Nevertheless, we performed a more detailed analysis of the adversarial users' characteristics and behaviors in~\cite{adversarial}.
Our manual validation there shows that the most active users are unlikely to be entirely automated by algorithms.
We also compared our data with a list of state controlled accounts with over 5,000 followers published by Twitter~\cite{twitter2018midterm}~(as the information of this set of accounts was not anonymized):~none of them showed up in our dataset.

\subsubsection{Political preference labeling evaluation.} 
To validate labels on user political preference,
we sample 100 labeled users and at most $10$ tweets 
from each user. 
We ask raters to label each user as pro-Democrat, pro-Republican or unclear
given the tweets and users' profiles.
Raters agree with the machine-assigned
labels for $93\%$ of the users.

\subsubsection{Evaluating DPP.}
With inferred user political preference, we can now apply the DPP method
to identify if a given tweet is adversarial towards a candidate.
We evaluate our method on the sampled datasets in Table~\ref{tab:directedSample}.
For some of the tweets, we are unable to collect enough user information for political preference labeling or the content is too short to be scored by Perspective API.
Such tweets comprise $8\%$ of all interactions in our dataset. 
In the middle section of the table, we show the number of tweet-candidate pairs
that we have enough information to label with DPP,
along with the number of adversarial tweets that are directed at the candidate in the labeled set.
Compare to the results as shown in the second row,
the ratio of directed adversarial tweets in the DPP-labeled set remains unchanged.

As our goal is to have a high-precision and scalable approach to quantify adversarial interactions received per candidates,
we focus on measuring the precision of DPP in identifying directed adversarial content.
For a given candidate-tweet pair,
a positive label is when the tweet is labeled as containing adversarial content against the candidate.
The evaluation was performed on the subset of the tweet-candidate pairs that are automatically labeled by DPP.
We define the DPP precision as 
the percentage of true positive labels over all the tweet-candidate pairs that are machine-labeled as positive.
Likewise, DPP recall is defined as the percentage of true positive labels over all the positive tweet-candidate pairs that are manually labeled as positive. 
In the bottom section of Table~\ref{tab:directedSample},
we show the precision and recall of DPP. We emphasize that this is only recall
measured relative to the set of DPP-labeled tweets, not to the overall set of tweets in our dataset.
Since the ratio of directed adversarial tweets remains unchanged
in the DPP-labeled dataset as listed in the two middle rows in the table, the baseline approach of using Perspective
API with Twitter metadata would result in the precision as shown in the second
row (e.g., 75\% for all replies).
Thus our method improves over
the baseline approach across
candidates with different popularity levels
while achieving over $80\%$ recall in most cases.

For all candidates, on average, 
the measured quantity of
adversarial interactions decrease by $36\%$ when applying
DPP compared to using the approach of combining
Perspective API and Twitter metadata.  For some candidates such as Janet Garrett
(running against Jim Jordan), Andrew Janz (against Devin Nunes), Omar Navarro
(against Maxine Waters), the quantity of adversarial interactions decreases from more than $9\%$ of all interactions towards them to less than
$1.5\%$.
The opponents of these three candidates attract significant
attention, hence a large proportion of adversarial tweets interacting with them target their opponents.

\section{Measuring Adversarial Interactions \\ Towards Candidates}

With DPP, we can now quantify the scale of adversarial
interactions against candidates and compare the quantity of adversarial interactions
based on candidates' characteristics, including demographics or party affiliation.

In the following regression analysis, we show the impact of various candidate attributes on the dependent variable -- amount of directed adversarial interactions received per candidate estimated by DPP.
The model controls for the number of followers a candidate has and the amount of attention the candidate received from users who are in favor of the opposing party of them.
Candidate gender and affiliated party are represented as binary features.
Variable base levels~(zero) are Democrats for party affiliation and female for gender.
We experimented with number of candidate posts and overall attention towards a candidate as independent variables and excluded them because of high collinearity with the total amount of interaction a candidate attracts from users supporting the opposing party.

For the regression, we excluded the top and bottom $5\%$ of candidates in our dataset based on the overall attention they received.
We removed the top candidates since these candidates receive national attention, 
resulting in entirely different interaction dynamic and content than those seen with other candidates.
We exclude the bottom-attention candidates as they do not get any attention at all (less than seven tweets on average during that period).
We ran the same regression model excluding a larger set of $10\%$, $15\%$ and $20\%$ of candidates.
The results are highly similar across all these levels, while significantly different from analysis including all candidates.

Table~\ref{tab:regression} includes the details of the regression analysis.
Histograms in the table show the distribution of variables, along with the mean and standard deviation~(rightmost three columns).
We log-transformed 
the number of followers and amount of attention from users supporting the opposing party~(listed in italic in Table~\ref{tab:regression}),
so that values are normally distributed as shown in the histograms.
Further, as suggested in previous work~\cite{gelman2008scaling},
all continuous
variables were standardized by centering and dividing by two standard deviations in order to have the same scale for comparison with the binary variables, which were centered.  
The standardized beta coefficients ($B$) and their standard errors ($SE$) of both models are also presented in Table~\ref{tab:regression}. 
The $p$ values are computed from two-tailed T test.
For independent variables included in both models, variance inflation factors~(VIFs) are less than $2$, indicating
that multicollinearity is not an issue.
As shown in the bottom row of the table,
our model explains
$R^2 = 0.388$ of the variance of the dependent
variable.

Among the variables used in the model, we can see that the most significant predictor of adversarial interactions against a candidate is the amount of attention they receive from users who are in favor of the opposing party~(this variable was highly correlated with overall attention to the candidate, not used in the model, as noted above).
In the model, ten times more interactions a candidate gets from supporters of the opposing party adds $88.7$ more adversarial interactions~($117.1$ divided by twice the standard deviation, $0.66$).
We also notice that the number of followers negatively correlates with the dependent variable, ten times more followers a candidate gets decreases the number of adversarial interactions by $25$,
when holding the other variables constant.
Other variables are not as predictive on the quantity of estimated adversarial interactions received per candidate.
In our dataset, the majority of all users---$68\%$---are estimated as pro-Democrats.
When this skewed distribution of party preference of users on Twitter is controlled via the amount of attention coming from opponent users in the regression analysis,
the gender and affiliated party of a candidate are not predictive of the quantity of adversarial interactions.

In conclusion, our method for directed detection of adversarial interactions allows quantifying and comparing adversarial interactions targeting candidates. Our findings suggest that the overall attention to the candidates is the main predictor of adversarial interactions, and that party or gender, for example, are not significant factors for the candidates in our (trimmed) dataset.

\section{Discovering Candidate-Specific\\ Adversarial Interactions}

The analysis above focused on quantifying adversarial interactions that contain toxic speech. 
However, as noted above, adversarial interactions 
can often be very subtle and do not necessarily contain language that can be easily
flagged by context-agnostic language-based methods.
To better understand the diversity of adversarial interactions,
and to discover content that is missed by the methods based on context-agnostic models,
we develop an algorithm
that can discover \textit{target-specific adversarial lexicons}. 
Specifically, the algorithm assigns 
\textit{Adversary Scores} to terms used in interactions with a specific candidate, in 
order to discover terms that are frequently used in an adversarial way towards the candidate. 

Our approach builds on previous work which identifies domain-specific sentiment terms using a graph-based label propagation approach called \textsc{SentProp}~\cite{hamilton2016inducing}.
Specifically, \textsc{SentProp} computes embeddings for each word from their co-occurrences on a domain-specific corpus and 
connects words that are semantically close based on the cosine similarities of the respective word vectors. 
After the graph is constructed, a subset of words is initialized from a seed set as having positive or negative polarisation. The algorithm then propagates the positive and negative polarity scores from the seed sets by performing random walks on the graph. Finally, \textsc{SentProp} selects sentiment lexicons given the resulting polarity scores.

We adapt the \textsc{SentProp} framework in several key ways to induce terms likely to be used in an adversarial manner towards individual candidates. 
Broadly, we construct an interaction graph for each individual candidate. 
Since the likelihood of a term being used in an adversarial manner also depends on the sender of the message, we construct a joint graph of users and terms interacting with the candidate.
Finally, we initialize the node scores in the graph using a seed set of \textit{users} that used explicit adversarial language towards any candidate.  

\subsubsection{Constructing a user-term graph.} 

More formally, we create a candidate-specific corpus $C_P$ for each candidate
$P$ comprising all tweets interacting with the candidate, i.e., all tweets
replying to or mentioning the candidate in our data. We create the set $T$ of all unigram terms used by at
least $10$ users, excluding stopwords and candidate names, and the set $U$ of
users who used at least one term from $T$ while interacting with candidate $P$.
We then construct a user-term graph
$G_P = (V, E)$ where $V = T \cup U$. 
We connect the vertices with edges $E = E_U \cup E_{UT} \cup E_{T}$: a set of
connections between users ($E_U$, $\{v_{u_2} \to v_{u_1}$ for any user $u_1$
following $u_2\}$ on Twitter), between terms ($E_T$, $\{v_{t_1} \to v_{t_2}$ if
$t_2$ is one of $t_1$'s $k$ nearest neighbours in the word embedding space$\}$,
using $k=10$), and between users and terms ($E_{UT}$)  if
the user uses the term in interactions towards the candidate $P$ ($\{v_{u}
\leftrightarrow v_{t}$ for any user who used term $t\}$). 

\begin{table*}[t]
  \centering
  \small
  \begin{tabular}{p{20mm}lrp{88 mm}r}
  \toprule
  Candidate          & Term & Adversary score  & Sample tweet & \% of ``toxic'' \\
  & &  (confidence) & & tweets \\
  \midrule
  Ammar Campa-Najjar & terrorist & $0.73$ $(\pm 0.06)$ &  @ACampaNajjar Isn't you grand father a high ranking \textbf{terrorist} or Taliban member? & $35\%$\\
  
  Ilhan Omar & brother & $0.82$ $(\pm 0.01)$ & @IlhanMN did you and your \textbf{brother} have fun on your honeymoon?? & $21\%$\\

  Duncan Hunter & wife & $0.99$ $(\pm 0.01)$ & @Rep\_Hunter Hey Duncan baby how is your \textbf{wife}?  Getting better after being thrown under the bus? & $34\%$ \\

  Lena Epstein & rabbi & $0.99$ $(\pm 0.00)$ & @LenaEpstein Are you a stupid person, you had a Christian \textbf{rabbi} to your event. You suck at being a Jew, you and Jared Kushner are the worst. & $18\%$ \\

  \bottomrule
  \end{tabular}
  \caption{\label{tab:attack} Examples of candidate-specific adversarial terms picked up by our method, along with sample tweets containing these terms. In addition, we show the proportion of tweets being labeled as ``toxic'' by Perspective API, among all interactions with the candidate containing the term while coming from a user in favor of the opposing party.}
\end{table*}
\normalsize

The edge weights among terms, i.e., edges in $E_T$, are set to the cosine-similarity between their term embeddings, normalized such that the maximum edge weight is 1. The term embeddings are trained on corpus $C_P$ following the method used in \textsc{SentProp}~\cite{hamilton2016inducing}.
Edge weights in $E_{UT}$ are set as the frequency of user $u$ using term $t$ normalized by the frequency of the user interacting with candidate $P$. Finally, edge weights in $E_U$ are set to $1$.

\subsubsection{Propagating adversary scores from a seed set of users.}
We initialize adversary scores for a seed set of users who have posted adversarial interactions to any candidate. 
We identify such users by taking into account their party preference and whether they performed explicit~(i.e., detected by context-agnostic methods) toxic interactions towards candidates of the opposing party. Specifically, we construct two overall seed sets: $U_D$ is a set of all pro-Democrat users in the dataset who have toxic interactions with any Republican candidate, i.e., interactions labeled with a toxicity score larger than 0.7 by Perspective API, and analogously a set of pro-Republican users $U_R$.

Then, for each candidate, we initialize adversary scores as $1.0$ for the seed
set of users in favor of the candidate's opposing party, e.g., for Republican
candidate $P$ we set $U_{P_{seed}} = U_P \cap U_D$. After the seed set is
initialized, we propagate the adversary scores over the graph using a random
walk approach~\cite{zhou2004learning}. A term's adversary score towards the candidate is set as the probability of a random walk from the seed set reaching that node, normalized so that the maximum value is $1$.
Finally, as in~\cite{hamilton2016inducing}, to ensure robustness of adversary scores, we perform label propagation for each candidate 50 times using a random selection of $70\%$ of the seed nodes each time. The confidence of term adversary score is defined as the standard deviation of the computed scores across runs. For most terms, the adversary score remains stable.

\subsubsection{Discovery of adversarial terms.} 
We run the analysis for the 235 candidates who received at least $800$ replies or mentions in our data.
For the evaluation, we select the 50 term-candidate pairs with the highest adversary score that result from these analyses. For each term-candidate pair we sample 10 tweets that match  the pair in order to examine how these terms are used against these candidates.

Our evaluation shows that a majority of the sampled terms are indeed used adversarially,
with almost half~($21$ out of $50$) of the terms unlikely to be captured by context-agnostic models.
Specifically, our evaluation found that in $15$ of the $50$ cases it is hard to associate the discovered terms with one single topic, like ``vote'' or ``people''.  
We found 14 terms that are explicitly adversarial, like ``lie'' or ``racist''. 
Finally, $21$ of the $50$ terms, in our evaluation, exposed adversarial interactions that used novel language. 
One example is ``Fartenberry'', from the graph for congressman Jeff Fortenberry, a derogatory name used by his opponents. 
Another discovered term, ``family'', came from the graph for Republican representative Paul Gosar, used by opponents to mock the candidate whose six siblings endorsed his Democrat opponent.

We show selected samples of candidate-term pairs with high adversary scores and a tweet containing the term in Table~\ref{tab:attack}.
The table shows the percentage of the tweets of each pair -- that is all interactions with the candidate containing the term and posted by users in favor of the opponent party of the candidate --  that were labeled as toxic by Perspective API.
The results show that such content is largely undetected by Perspective API. 
For example, Ammar Campa-Najjar,
whose grandfather was assassinated due to the suspicion that he masterminded the 
1972 Munich massacre, was accused of being a terrorist himself.
In total, we found $51$ tweets from pro-Republican users interacting with Campa-Najjar,
while referring to the accusation using the term ``terrorist'', which is not generally highligheted as adversarial by context-agnostic approaches.
Ilhan Omar, who is falsely accused of marrying her brother for him to gain
permanent residency in the US,
received $695$ tweets from pro-Republican users with term ``brother'',
referring to the alleged incident.

Our approach captures the cases of using misinformation to undermine the legitimacy of candidates, but adversarial interactions were not limited to misinformation.
For instance, as a consequence of blaming his wife for the charge of embezzlement,
Duncan Hunter received $314$ tweets with the term ``wife'' criticizing him over the issue. 
Similarly, 
because of inviting a Messianic rabbi to a rally after the Pittsburgh synagogue shooting,
of 1,945 tweets interacting with Lena Epstein in our data, $15\%$ criticized her for this incident (term: ``rabbi''). 

In conclusion, this informal evaluation shows that the algorithm we developed can discover adversarial interactions that are missed by context-agnostic methods.
In the next section, we use some of the 
tweets and terms from the sample set we collect, and combine them with previous work on online harassment to provide a typology of adversarial interactions against political candidates. 

\vfill\null

\section{Discussion: The Many Kinds \\of Adversarial Interactions}

Using the approach we developed, we were able to discover samples of more subtle adversarial interactions. 
In this section, we provide a typology of adversarial interactions against political candidate, using the types identified in our work here as well as earlier victim-reported harassment categories from previous work~\cite{duggan2014online,matias2015reporting}. 
Rather than offer an exhaustive taxonomy,
we hope to emphasize the challenges facing both accurately annotating and detecting adversarial interactions, by illustrating these types with examples.

\subsubsection{Offensive name-calling.}
Explicit insults and usage of abusive language are a common form of harassment on Twitter.
Examples often contain terms that are toxic independent of the specific victim.
Illustrative examples in our datset include \textit{``@RepMcCaul Stop using so much water you ass clown. We're having a water crisis.''} and \textit{``@VoteRobDavidson You are a joke. \#RadicalRob''}.
Similar utterances are likely to be perceived as insults in other online forums, where conversations are often used as training data for machine learning methods to detect toxicity~\cite{wulczyn2017ex}.
In consequence, adversarial interactions in this category can be detected relatively well via automatic methods. 
However, as we show above, offensive name calling in Twitter replies are not always directed towards the recipient, complicating the analysis.

\subsubsection{Threats of violence.}
Another prominent type of adversarial interaction are tweets that threaten to cause physical or mental pain. 
Often, these threats are rather explicit and thus relatively robust to detect via automated methods, such as \textit{``@RepSwalwell FU MR. Trump! You need someone to tie you down and torture and rape you. \#Deliverance''} and \textit{``@Jim\_Jordan You will burn in hell.''}

While some of these threats can be easily detected by context-agnostic models, others can be implicit and require context to interpret.
For example, a month before mailing $16$ bombs to several prominent critics of President Trump, 
Cesar Altieri Sayoc Jr. sent the following tweets to Senator Elizabeth Warren: 
\textit{``@SenWarren @SecretaryZinke A Promise we will see you real soon. Hug
loved one everytime you leave your home''} and \textit{``@SenWarren @SecretaryZinke Nice home in Cambridge see you soon''}.
Although none of the bombs were sent to Senator Warren, given the context, these tweets are likely to have been threats.
Without hindsight and knowing the identity of the sender of these messages,
they can be interpreted in many different ways. This use of language poses significant challenges towards automated threat detection.

\subsubsection{Posting discrediting information.}
A common adversarial tactic in our data involves spreading information with the aim of discrediting the candidate. 
This can include both adversarially posting misinformation and sharing true information about a candidate in a hostile way.
Alleged scandals involving candidates, whether or not they are true,
are often used in an adversarial manner.
For example, Ilhan Omar has been falsely accused of marrying her brother.
Many tweets in our dataset referred to this claim, some more explicitly, e.g. \textit{``Weren't you the one who married your Brother?''}, 
and others more implicitly, e.g.  \textit{``Will your brother be there?''}. 

While we discovered these tweets using our tool for target-specific adversarial lexicons detection, 
this category of adversarial interaction is hard to be accurately
detected, even by humans.
Since these attacks are usually tailored to a specific individual,
their detection and interpretation often requires background knowledge that's only known to a small group of people.
Even more difficult is differentiating  misinformation from hostile but true information~(e.g. scandals of political candidates).
While new approaches are developed to allow for context in interpreting and labeling social data~\cite{patton2020vatas}, this remains a challenging and time-consuming task.

\subsubsection{Attacks on identity.}
Attacks on the basis of attributes such as race, religion, or gender identity are common.
Examples include misogynist speech, such as \textit{``@Ocasio2018 You sexy little tart\ldots I don't do socialism, but I will do you, you hot little Latina socialist, as long as you don't talk politics and you do make me a sandwich afterward\ldots''},
hate speech targeting minority groups like \textit{``@RepMaxineWaters Hey Maxine don’t monkey this thing up please''}, and other identity-based attacks.

Correct interpretation of these insults also requires understanding of the context, hence making it hard for raters and automated detection.
For example, complementing one's appearance is generally not offensive,
however repeatedly calling a female politician ``hot'' in a political discussion
is considered inappropriate, at least in the United States. 

\subsubsection{Adversarial message repetition.}
Message repetition is an effective way to sway an audience~\cite{cacioppo1979effects}.
We observe cases where adversarial messages or topics that are repeated
to amplify their impact in terms of discouraging candidates.
Examples include repeating negative messages,
like the user who sent 18 tweets of \textit{``this is disgusting,  a horror.  Resign!''} to Nancy Pelosi, 
or misleading information, such as made-up scandals.

Classifying this type of adversarial interaction faces significant challenges.
First, messages expressing legit political request might as well be repeated multiple times. 
A careful definition is required to distinguish adversarial repetition from practise of democracy.
Complicating further, identification of adversarial message repetition requires human raters to
annotate multiple utterances,
instead of one utterance alone as is common for most rating tasks.

\section{Conclusion and Future Work}
In this work, we analyzed adversarial interactions on Twitter towards political candidates 
during the run-up to the US 2018 midterm general election.
We collected and analyzed a dataset of 1.7 million tweets.
We leveraged the bipartisan nature of US politics to design a method that combines
heuristics and context-agnostic language-based algorithms in order to infer the
targets of adversarial content on Twitter.
This method allows us to better quantify adversarial interactions towards candidates.
Our findings show that the overall attention a candidate gets on Twitter is the most indicative factor of adversarial interactions they receives.
Other candidate attributes such as affiliated party or gender have no significant impact when predicting quantity of adversarial interactions against a candidate.

While the method we used achieved high precision, we further explored the type of interactions that are missed by context-agnostic models.
To this end, we developed a novel algorithm to discover \textit{target-specific
adversarial lexicons} by combining language and social network features.
Our approach exposes adversarial
interactions that are not readily discovered by context-agnostic generalized
language-based tools. Combining our discovery with victim-reported harassment
types, we highlighted the fact that adversarial interactions remains a challenging
task for both machine and human raters, as many types of adversarial interactions require context and background knowledge to interpret.
Although our method is mainly designed for adversarial interaction detection, we believe that this ``personalized'' approach for detecting challenging language directed at a single individual could be useful in other tasks, including hate speech or misinformation detection.

Interactions with political figures and candidates on Twitter (as well as on other public forums) are potentially different from the interactions with other public figures.
In the US, for example, the courts have been debating the right of public officials to block a person from interacting with their Twitter account~\cite{knight}, as such behavior may violate the person's First Amendment~(i.e. free speech) rights.
Regardless of the courts' final decision, platforms like Twitter may use the methods we propose here to improve the detection of adversarial interactions with political figures.
Such detection can foster better discourse around political figures, for example by ranking and demoting content,
or by better identifying users who are instigators of toxic campaigns and environment~\cite{adversarial}, and devising measures to handle such offenders.

\section{Acknowledgements}

We thank Jingxuan Sun, Xiran Sun and Julia Narakornpichit for data labeling support.
We thank Andreas Veit and the anonymous reviewers for their feedback.
This research is supported by NSF research grants CNS-1704527 and IIS-1665169, as well as a Cornell Tech Digital Life Initiative Doctoral Fellowship.

\bibliography{citation}

\begin{thebibliography}{}

\bibitem[\protect\citeauthoryear{{Amnesty International UK}}{2018}]{ukreport}
{Amnesty International UK}.
\newblock 2018.
\newblock Black and asian women mps abused more online.
\newblock \url{https://www.amnesty.org.uk/online-violence-women-mps}.
\newblock [Online; accessed 5-Jan-2019].

\bibitem[\protect\citeauthoryear{Arif, Stewart, and
  Starbird}{2018}]{arif2018acting}
Arif, A.; Stewart, L.~G.; and Starbird, K.
\newblock 2018.
\newblock Acting the part: Examining information operations within
  \#blacklivesmatter discourse.
\newblock ACM.

\bibitem[\protect\citeauthoryear{Astor}{2018}]{nytimes}
Astor, M.
\newblock 2018.
\newblock For female candidates, harassment and threats come every day.
\newblock
  \url{https://www.nytimes.com/2018/08/24/us/politics/women-harassment-elections.html}.
\newblock [Online; accessed 5-Jan-2019].

\bibitem[\protect\citeauthoryear{Ballotpedia}{2018}]{ballotpedia}
Ballotpedia.
\newblock 2018.
\newblock List of candidates who ran in u.s. congress elections, 2018.
\newblock
  \url{https://ballotpedia.org/List_of_candidates_who_ran_in_U.S._Congress_elections,_2018}.
\newblock [Online; accessed 15-Jan-2019].

\bibitem[\protect\citeauthoryear{Barber{\'a}}{2015}]{barbera2015birds}
Barber{\'a}, P.
\newblock 2015.
\newblock Birds of the same feather tweet together: Bayesian ideal point
  estimation using twitter data.
\newblock {\em Political Analysis} 23(1):76--91.

\bibitem[\protect\citeauthoryear{Buchwald}{2018}]{knight}
Buchwald, N.~R.
\newblock 2018.
\newblock Knight v. trump memorandum and order.
\newblock \url{https://bit.ly/2LDgTPL}.
\newblock [Online; accessed 11-May-2019].

\bibitem[\protect\citeauthoryear{Cacioppo and
  Petty}{1979}]{cacioppo1979effects}
Cacioppo, J.~T., and Petty, R.~E.
\newblock 1979.
\newblock Effects of message repetition and position on cognitive response,
  recall, and persuasion.
\newblock {\em Journal of Personality and Social Psychology} 37(1):97.

\bibitem[\protect\citeauthoryear{Chatzakou \bgroup et al\mbox.\egroup
  }{2017a}]{chatzakou2017hate}
Chatzakou, D.; Kourtellis, N.; Blackburn, J.; De~Cristofaro, E.; Stringhini,
  G.; and Vakali, A.
\newblock 2017a.
\newblock Hate is not binary: Studying abusive behavior of\# gamergate on
  twitter.
\newblock In {\em Proceedings of the 28th ACM Conference on Hypertext and
  Social Media},  65--74.
\newblock ACM.

\bibitem[\protect\citeauthoryear{Chatzakou \bgroup et al\mbox.\egroup
  }{2017b}]{chatzakou2017mean}
Chatzakou, D.; Kourtellis, N.; Blackburn, J.; De~Cristofaro, E.; Stringhini,
  G.; and Vakali, A.
\newblock 2017b.
\newblock Mean birds: Detecting aggression and bullying on twitter.
\newblock In {\em Proceedings of the 2017 ACM on Web Science Conference},
  13--22.
\newblock ACM.

\bibitem[\protect\citeauthoryear{Chatzakou \bgroup et al\mbox.\egroup
  }{2017c}]{chatzakou2017measuring}
Chatzakou, D.; Kourtellis, N.; Blackburn, J.; De~Cristofaro, E.; Stringhini,
  G.; and Vakali, A.
\newblock 2017c.
\newblock Measuring\# gamergate: a tale of hate, sexism, and bullying.
\newblock In {\em Proceedings of the 26th international conference on World
  Wide Web companion},  1285--1290.
\newblock International World Wide Web Conferences Steering Committee.

\bibitem[\protect\citeauthoryear{Conover \bgroup et al\mbox.\egroup
  }{2011}]{conover2011political}
Conover, M.; Ratkiewicz, J.; Francisco, M.~R.; Gon{\c{c}}alves, B.; Menczer,
  F.; and Flammini, A.
\newblock 2011.
\newblock Political polarization on twitter.
\newblock In {\em Proceedings of the International AAAI Conference on Web and
  Social Media}.
\newblock AAAI Press.

\bibitem[\protect\citeauthoryear{Davis \bgroup et al\mbox.\egroup
  }{2016}]{davis2016botornot}
Davis, C.~A.; Varol, O.; Ferrara, E.; Flamini, A.; and Menczer, F.
\newblock 2016.
\newblock Botornot: A system to evaluate social bots.
\newblock In {\em Proceedings of the 25th International Conference on World
  Wide Web}.
\newblock International World Wide Web Conferences Steering Committee.

\bibitem[\protect\citeauthoryear{Duggan}{2017}]{duggan2014online}
Duggan, M.
\newblock 2017.
\newblock {\em Online harassment}.
\newblock Pew Research Center.

\bibitem[\protect\citeauthoryear{ElSherief \bgroup et al\mbox.\egroup
  }{2018}]{elsherief2018hate}
ElSherief, M.; Kulkarni, V.; Nguyen, D.; Wang, W.~Y.; and Belding, E.
\newblock 2018.
\newblock Hate lingo: A target-based linguistic analysis of hate speech in
  social media.
\newblock In {\em Proceedings of the International AAAI Conference on Web and
  Social Media}.
\newblock AAAI Press.

\bibitem[\protect\citeauthoryear{Ferrara \bgroup et al\mbox.\egroup
  }{2016}]{ferrara2016rise}
Ferrara, E.; Varol, O.; Davis, C.; Menczer, F.; and Flammini, A.
\newblock 2016.
\newblock The rise of social bots.
\newblock {\em Communications of the ACM} 59(7):96--104.

\bibitem[\protect\citeauthoryear{Finkelstein \bgroup et al\mbox.\egroup
  }{2018}]{finkelstein2018quantitative}
Finkelstein, J.; Zannettou, S.; Bradlyn, B.; and Blackburn, J.
\newblock 2018.
\newblock A quantitative approach to understanding online antisemitism.
\newblock {\em arXiv preprint arXiv:1809.01644}.

\bibitem[\protect\citeauthoryear{Founta \bgroup et al\mbox.\egroup
  }{2018}]{founta2018large}
Founta, A.~M.; Djouvas, C.; Chatzakou, D.; Leontiadis, I.; Blackburn, J.;
  Stringhini, G.; Vakali, A.; Sirivianos, M.; and Kourtellis, N.
\newblock 2018.
\newblock Large scale crowdsourcing and characterization of twitter abusive
  behavior.
\newblock In {\em Proceedings of the International AAAI Conference on Web and
  Social Media}.
\newblock AAAI Press.

\bibitem[\protect\citeauthoryear{Gelman}{2008}]{gelman2008scaling}
Gelman, A.
\newblock 2008.
\newblock Scaling regression inputs by dividing by two standard deviations.
\newblock {\em Statistics in Medicine} 27(15):2865--2873.

\bibitem[\protect\citeauthoryear{Gorrell \bgroup et al\mbox.\egroup
  }{2018a}]{gorrell2018online}
Gorrell, G.; Greenwood, M.; Roberts, I.; Maynard, D.; and Bontcheva, K.
\newblock 2018a.
\newblock Online abuse of uk mps in 2015 and 2017: Perpetrators, targets, and
  topics.
\newblock {\em arXiv preprint arXiv:1804.01498}.

\bibitem[\protect\citeauthoryear{Gorrell \bgroup et al\mbox.\egroup
  }{2018b}]{gorrell2018twits}
Gorrell, G.; Greenwood, M.~A.; Roberts, I.; Maynard, D.; and Bontcheva, K.
\newblock 2018b.
\newblock Twits, twats and twaddle: Trends in online abuse towards uk
  politicians.
\newblock In {\em Proceedings of the International AAAI Conference on Web and
  Social Media}.
\newblock AAAI Press.

\bibitem[\protect\citeauthoryear{Hamilton \bgroup et al\mbox.\egroup
  }{2016}]{hamilton2016inducing}
Hamilton, W.~L.; Clark, K.; Leskovec, J.; and Jurafsky, D.
\newblock 2016.
\newblock Inducing domain-specific sentiment lexicons from unlabeled corpora.
\newblock In {\em Proceedings of the Conference on Empirical Methods in Natural
  Language},  595.
\newblock Association for Computational Linguistics.

\bibitem[\protect\citeauthoryear{Hua, Naaman, and
  Ristenpart}{2020}]{adversarial}
Hua, Y.; Naaman, M.; and Ristenpart, T.
\newblock 2020.
\newblock Characterizing twitter users who engage in adversarial interactions
  against political candidates.

\bibitem[\protect\citeauthoryear{Hutto and Gilbert}{2014}]{hutto2014vader}
Hutto, C.~J., and Gilbert, E.
\newblock 2014.
\newblock Vader: A parsimonious rule-based model for sentiment analysis of
  social media text.
\newblock In {\em Proceedings of the International AAAI Conference on Web and
  Social Media}.
\newblock AAAI Press.

\bibitem[\protect\citeauthoryear{Im \bgroup et al\mbox.\egroup
  }{2019}]{im2019still}
Im, J.; Chandrasekharan, E.; Sargent, J.; Lighthammer, P.; Denby, T.; Bhargava,
  A.; Hemphill, L.; Jurgens, D.; and Gilbert, E.
\newblock 2019.
\newblock Still out there: Modeling and identifying russian troll accounts on
  twitter.
\newblock {\em arXiv preprint arXiv:1901.11162}.

\bibitem[\protect\citeauthoryear{Jigsaw}{2018}]{perspectiveapi}
Jigsaw.
\newblock 2018.
\newblock Perspective api.
\newblock \url{https://www.perspectiveapi.com/}.
\newblock [Online; accessed 3-Jan-2019].

\bibitem[\protect\citeauthoryear{Matias \bgroup et al\mbox.\egroup
  }{2015}]{matias2015reporting}
Matias, J.; Johnson, A.; Boesel, W.~E.; Keegan, B.; Friedman, J.; and DeTar, C.
\newblock 2015.
\newblock Reporting, reviewing, and responding to harassment on twitter.
\newblock {\em Available at SSRN 2602018}.

\bibitem[\protect\citeauthoryear{Mondal, Silva, and
  Benevenuto}{2017}]{mondal2017measurement}
Mondal, M.; Silva, L.~A.; and Benevenuto, F.
\newblock 2017.
\newblock A measurement study of hate speech in social media.
\newblock In {\em Proceedings of the 28th ACM Conference on Hypertext and
  Social Media},  85--94.
\newblock ACM.

\bibitem[\protect\citeauthoryear{Monje~Jr}{2019}]{twitter2018midterm}
Monje~Jr, C.
\newblock 2019.
\newblock 2018 us midterm elections review.
\newblock
  \url{https://blog.twitter.com/en_us/topics/company/2019/18_midterm_review.html}.
\newblock [Online; accessed 20-Aug-2019].

\bibitem[\protect\citeauthoryear{Nobata \bgroup et al\mbox.\egroup
  }{2016}]{nobata2016abusive}
Nobata, C.; Tetreault, J.; Thomas, A.; Mehdad, Y.; and Chang, Y.
\newblock 2016.
\newblock Abusive language detection in online user content.
\newblock In {\em Proceedings of the 25th International Conference on World
  Wide Web},  145--153.
\newblock International World Wide Web Conferences Steering Committee.

\bibitem[\protect\citeauthoryear{Patton \bgroup et al\mbox.\egroup
  }{2020}]{patton2020vatas}
Patton, D.~U.; Blandfort, P.; Frey, W.~R.; Schifanella, R.; McGregor, K.; and
  Chang, S.-F.~U.
\newblock 2020.
\newblock Vatas: An open-source web platform for visual and textual analysis of
  social media.
\newblock {\em Journal of the Society for Social Work and Research}
  11(1):000--000.

\bibitem[\protect\citeauthoryear{Raghavan, Albert, and
  Kumara}{2007}]{raghavan2007near}
Raghavan, U.~N.; Albert, R.; and Kumara, S.
\newblock 2007.
\newblock Near linear time algorithm to detect community structures in
  large-scale networks.
\newblock {\em Physical review E} 76(3):036106.

\bibitem[\protect\citeauthoryear{Ribeiro \bgroup et al\mbox.\egroup
  }{2018}]{ribeiro2018characterizing}
Ribeiro, M.~H.; Calais, P.~H.; Santos, Y.~A.; Almeida, V.~A.; and Meira~Jr, W.
\newblock 2018.
\newblock Characterizing and detecting hateful users on twitter.

\bibitem[\protect\citeauthoryear{Romero, Tan, and
  Ugander}{2013}]{romero2013interplay}
Romero, D.~M.; Tan, C.; and Ugander, J.
\newblock 2013.
\newblock On the interplay between social and topical structure.
\newblock In {\em Proceedings of the International AAAI Conference on Web and
  Social Media}.
\newblock AAAI Press.

\bibitem[\protect\citeauthoryear{Salminen \bgroup et al\mbox.\egroup
  }{2018}]{salminen2018anatomy}
Salminen, J.; Almerekhi, H.; Milenkovic, M.; Jung, S.-g.; An, J.; Kwak, H.; and
  Jansen, B.~J.
\newblock 2018.
\newblock Anatomy of online hate: Developing a taxonomy and machine learning
  models for identifying and classifying hate in online news media.
\newblock In {\em Proceedings of the International AAAI Conference on Web and
  Social Media}.
\newblock AAAI Press.

\bibitem[\protect\citeauthoryear{Theocharis \bgroup et al\mbox.\egroup
  }{2016}]{theocharis2016bad}
Theocharis, Y.; Barber{\'a}, P.; Fazekas, Z.; Popa, S.~A.; and Parnet, O.
\newblock 2016.
\newblock A bad workman blames his tweets: the consequences of citizens'
  uncivil twitter use when interacting with party candidates.
\newblock {\em Journal of communication} 66(6):1007--1031.

\bibitem[\protect\citeauthoryear{Twitter}{2019}]{twitterapi}
Twitter.
\newblock 2019.
\newblock Twitter standard api.
\newblock
  \url{https://developer.twitter.com/en/docs/tweets/filter-realtime/overview}.
\newblock [Online; accessed 15-Jan-2019].

\bibitem[\protect\citeauthoryear{Wulczyn, Thain, and
  Dixon}{2017}]{wulczyn2017ex}
Wulczyn, E.; Thain, N.; and Dixon, L.
\newblock 2017.
\newblock Ex machina: Personal attacks seen at scale.
\newblock In {\em Proceedings of the 26th International Conference on World
  Wide Web},  1391--1399.
\newblock International World Wide Web Conferences Steering Committee.

\bibitem[\protect\citeauthoryear{Yin \bgroup et al\mbox.\egroup
  }{2009}]{yin2009detection}
Yin, D.; Xue, Z.; Hong, L.; Davison, B.~D.; Kontostathis, A.; and Edwards, L.
\newblock 2009.
\newblock Detection of harassment on web 2.0.
\newblock In {\em Proceedings of the Content Analysis in the WEB}, volume~2,
  1--7.

\bibitem[\protect\citeauthoryear{Zhou \bgroup et al\mbox.\egroup
  }{2004}]{zhou2004learning}
Zhou, D.; Bousquet, O.; Lal, T.~N.; Weston, J.; and Sch{\"o}lkopf, B.
\newblock 2004.
\newblock Learning with local and global consistency.
\newblock In {\em Advances in Neural Information Processing Systems},
  321--328.
\newblock MIT Press.

\end{thebibliography}
\bibliographystyle{aaai}

\vfill

\section{Appendix I}

Full list of politically related hashtags in Table~\ref{tab:hashtags1}.

\begin{table}[h]
  \centering
  \begin{tabular}{p{15mm}p{60 mm}}
  \hline
  \texttt{\small{\#bluewave}} & \tiny{\texttt{\#glovesoff \#wherearethechildren \#lgbt \#liberal \#stillwithher \#traitortrump \#fbrparty \#resister \#resistance \#progressive \#flipitblue \#strongertogether \#trumprussia \#guncontrolnow \#fucktrump \#neveragain \#gunsense \#nevertrump \#metoo \#gunreformnow \#enough \#followbackresistance \#gunreform \#muellertime \#votethemout \#demforce \#uniteblue \#pru \#impeach45 \#voteblue \#bluetsunami \#feminist \#notmypresident \#blm \#persist \#bluewave \#antitrump \#dumptrump \#impeachtrump \#protectmueller \#familiesbelongtogether \#timesup \#theresistance \#lgbtq \#fbr \#democrat \#bluewave2018 \#climatechange \#takeaknee \#trumptreason \#blacklivesmatter \#basta \#marchforourlives \#resist \#boycottnra \#daca \#impeach \#guncontrol}}\\
  \texttt{\small{Filtered}} & \tiny{\texttt{\#atheist \#science \#equality \#vote  \#humanrights \#indivisible}} \\
  \hline
  \texttt{\small{\#maga}}     & \tiny{\texttt{\#lockherup \#tcot \#conservative  \#nra \#thegreatawakening \#2a \#draintheswamp \#trump2020 \#wwg1wga \#1a \#bluelivesmatter \#wethepeople \#maga \#kag2020 \#prolife \#buildthewall \#ccot \#fbts \#americafirst \#codeofvets \#trump \#backtheblue \#defundpp \#winning \#deplorable \#magaveteran \#redwave \#trumptrain \#kag \#nodaca \#potus \#covfefe \#greatawakening \#qanon \#walkaway \#molonlabe \#redwaverising \#termlimits}}\\
  \texttt{\small{Filtered}} & \tiny{\texttt{\#godblessamerica \#constitution \#military \#vets \#freedom \#usmc \#christian \#veterans \#veteran \#usa \#god \#codeofvets \#vet \#israel \#family \#jesus \#patriot \#america}} \\
  \hline
  \end{tabular}
  \caption{\label{tab:hashtags1} Hashtags related to \texttt{\#bluewave} or \texttt{\#maga} with filtered ones.}
\end{table}

\end{document}